\documentclass[aps,preprint,jcp, superscriptaddress, footinbib]{revtex4-1}
\usepackage{amssymb}
\usepackage{MnSymbol}
\usepackage{bm}% bold math
\usepackage{hyperref}% add hypertext capabilities
\usepackage{xcolor}

\usepackage{verbatim} 
\usepackage{calligra}

\usepackage{graphicx}% Include figure files
\usepackage[caption=false]{subfig} %multi figures
\graphicspath{{../forplot/}}

\newcommand{\Syn}{\mathsf{S}}
\newcommand{\PSD}{\mathsf{P}}
\newcommand{\Vol}{V} % system volume
\newcommand{\eng}{\epsilon_{\rm b}} % complex formation free energy
\newcommand{\engc}{\epsilon_{\rm ps}} % phase separation contact free energy
\newcommand{\kB}{k_{\rm B}}
\newcommand{\Zall}{{\cal Z}} 
\newcommand{\Fall}{{\cal F}}
\newcommand{\Kdr}{{\cal K}} % rescaled Kd
\newcommand{\psib}{\overline{\psi}} % mean-field psi

\begin{document}

%\title{Phase Separation of the SynGAP/PSD-95 Condensate Model of\\ 
%Postsynaptic Densities Involves Multivalent Interactions\\ 
%Auxiliary to Dilute-Phase Stoichiometric Binding} 

\title{Assembly of Model Postsynaptic Densities Involves Interactions\\ 
Auxiliary to Stoichiometric Binding}

\author{Yi-Hsuan Lin}
\thanks{Contributed equally to this work}
\address{Department of Biochemistry, University of Toronto, Toronto, Ontario M5S 1A8, Canada}
\address{Molecular Medicine, The Hospital for Sick Children, Toronto, Ontario
M5G 0A4, Canada}
\author{Haowei Wu}
\thanks{Contributed equally to this work}
\address{Division of Life Science, State Key Laboratory of Molecular 
Neuroscience, Hong Kong University of Science and Technology, 
Clear Water Bay, Hong Kong, China}
\author{Bowen Jia}
\address{Division of Life Science, State Key Laboratory of Molecular 
Neuroscience, Hong Kong University of Science and Technology, 
Clear Water Bay, Hong Kong, China}
\author{Mingjie Zhang}
\thanks{{\rm To whom correspondence should be addressed; Email:\\
{\tt zhangmj@sustech.edu.cn} (M.Z.);\\
{\tt huesun.chan@utoronto.ca} (H.S.C.)}}
\address{Division of Life Science, State Key Laboratory of Molecular 
Neuroscience, Hong Kong University of Science and Technology, 
Clear Water Bay, Hong Kong, China}
\address{School of Life Sciences, Southern University of Science and
Technology, Shenzhen, China}
\author{Hue Sun Chan}
\thanks{{\rm To whom correspondence should be addressed; Email:\\
{\tt zhangmj@sustech.edu.cn} (M.Z.);\\
{\tt huesun.chan@utoronto.ca} (H.S.C.)}}
\address{Department of Biochemistry, University of Toronto, Toronto, 
Ontario M5S 1A8, Canada}

\date{\today}	

\vfill\eject

\begin{abstract}

The assembly of functional biomolecular condensates often involves
liquid-liquid phase separation (LLPS) of proteins with multiple modular
domains, which can be folded or conformationally disordered to various
degrees. To understand the LLPS-driving domain-domain interactions,
a fundamental question is how readily the interactions in the condensed 
phase can be inferred from inter-domain interactions in dilute solutions. 
In particular, are the interactions leading to LLPS exclusively those 
underlying the formation of discrete inter-domain complexes in homogeneous 
solutions? We address this question by developing a mean-field LLPS 
theory of two stoichiometrically constrained solute species.  The theory 
is applied to the neuronal proteins SynGAP and PSD-95, whose complex
coacervate serves as a rudimentary model for neuronal postsynaptic
densities (PSDs). The predicted phase behaviors are compared with 
experiments. Previously, a three-SynGAP, two-PSD-95 ratio was 
determined for SynGAP/PSD-95 complexes in dilute solutions. However, when this 
3:2 stoichiometry is uniformly imposed in our theory encompassing both dilute 
and condensed phases, the tie-line pattern of the predicted SynGAP/PSD-95 phase
diagram differs drastically from that obtained experimentally. In contrast,
theories embodying alternate scenarios postulating auxiliary SynGAP-PSD-95 as
well as SynGAP-SynGAP and PSD-95-PSD-95 interactions in addition to those
responsible for stoichiometric SynGAP/PSD-95 complexes produce tie-line
patterns consistent with experiment. Hence, our combined 
theoretical-experimental analysis indicates that weaker interactions or 
higher-order complexes beyond the 3:2 stoichiometry, but not yet documented, 
are involved in the formation of SynGAP/PSD-95 condensates, imploring
future efforts to ascertain the nature of these auxiliary interactions in
PSD-like LLPS
and underscoring a likely general synergy between stoichiometric, structurally 
specific binding and stochastic, multivalent ``fuzzy'' interactions
in the assembly of functional biomolecular condensates.
\\

\end{abstract}

\noindent
{\bf Short Title:}\\ Interactions in SynGAP/PSD-95 Condensate
\\

\maketitle

\noindent
{\large\bf Statement of Significance}\\
$\null$\\
It has become increasingly clear that functional biomolecular condensates
underpinned by liquid-liquid phase separation (LLPS) are stabilized by dynamic
multivalent interactions as well as by structurally specific interactions. To
gain insights into the role of stoichiometric binding in biomolecular LLPS, we
develop theories for LLPS driven solely by interactions that stabilize
stoichiometric protein complexes in dilute solution and for alternate scenarios
in which auxiliary interactions also contribute. Application of our
formulations to experimental measurements of dilute- and condensed-phase
protein concentrations of the SynGAP/PSD-95 condensate model of postsynaptic
densities reveals that its assembly involves interactions auxiliary to those
stabilizing the 3:2 SynGAP/PSD-95 complex in dilute solution, exemplifying a
synergy between specific and stochastic interactions in the assembly of
biomolecular condensates. 
\\

\vfill\eject

\noindent
{\bf INTRODUCTION}

%%%%%%%%%%%%%%%%%%%%%%%%%%%%%%%%%%%%%%%%%%%%%%%%%%%%%%%%%%%%%%%%%%%%%%%%%%%%%%

%%%%%%%%%%%%%%%%%%%%%%%%%%%%%%%%%%%%%%%%%%%%%%%%%%%%%%%%%%%%%%%%%%%%%%%%%%%%%%

Biomolecular condensates composing of various proteins, nucleic acids, 
and small molecules are important for biological functions~\cite{Banani2017}. 
Examples include intracellular compartments, often referred
to as ``membraneless organelles'', e.g., P-granules, nuclear speckles, 
and Cajal bodies~\cite{Cliff2009, Spector2011}, and extracellular materials 
such as the precursory tropolelastin coacervates of elastic connective 
tissues~\cite{Fred2018}. Some membraneless organelles, such as nucleoli 
and stress granules, are organized further into 
subcompartments~\cite{Feric2016,JoshCliff2020}. Biomolecular 
condensates serve diverse functions, including but not limited to 
gene regulation, cell growth, and synaptic 
activities~\cite{Chong2016, Banani2017, Zeng2018a, Kim2019}. 
Membraneless organelles reduce biomolecular noise in the 
cell~\cite{julicher2020} and provide regulated 
local environments that facilitate specific biochemical 
processes, yet they can also assemble/disassemble rapidly in response to 
environmental changes~\cite{MitreaKriwacki2016}. 
Biophysically, liquid-liquid phase separation (LLPS) is recognized 
as a major---albeit not the only~\cite{Tjian2019}---mechanism in 
the assembly, including subcompartmentalization~\cite{Feric2016}, of 
many of these condensates~\cite{Tanja2019}.

Main ingredients of biomolecular condensates can be intrinsically disordered 
proteins (IDPs), folded globular proteins, RNA, and/or proteins with 
folded domains and intrinsically disordered regions 
(IDRs)~\cite{Nott2015,Feric2016,Deniz2017,RolandRev}. When only parts of 
the proteins (e.g., folded domains~\cite{Li2012}) engage in significant 
favorable interchain interactions whereas other parts (e.g., parts of or 
entire IDRs) do not, LLPS energetics may be conceptualized using a 
``stickers and linkers'' picture~\cite{Harmon2017,Harmon2018,lassi}. The 
sticker/linker distinction is not always clear-cut, however,
because every part of a protein chain may contribute to 
LLPS-driving interactions. For folded domains, the interactions can 
be---though not always---structurally specific, or entail
``folding upon binding'' when IDR interacting partners are
involved~\cite{uversky2002,cosb15}. In contrast, for interactions among 
IDPs/IDRs, the molecular recognition mechanisms underlying their
sequence-dependent LLPS~\cite{Nott2015,Lin16,JeetainPNAS2020,RohitSci2020,SumanPNAS2020}
are stochastic or ``fuzzy''~\cite{Fuxreiter2015} in that they involve diverse,
dynamic conformations and transient 
interactions~\cite{Lewis2017,Lin2017c,Alan2020,Paletal2021}.
However, some of the transient interactions may entail fluctuations between
disordered chain configurations with labile but nonetheless specific 
fibril-like structures~\cite{McKnight2017,McKnightRev2018}.

It has been known for some time that 
proteins participating in signaling pathways are often organized structurally 
and sequentially in a modular fashion, with folded domains such as SH2 and
SH3 acting as modules playing key regulatory roles~\cite{TonyPawson}.
Recently, multivalent interactions involving some of these domains are 
found to be important for the assembly of biomolecular condensates.
Thus, LLPS can be a key physico-chemical mechanism exploited by Nature
for physiological regulation as LLPS's essentially all-or-none features 
entail sharp transitions among conformationally and hence functionally 
distinct states of these proteins~\cite{Li2012}.

In view of the central role of multivalency in biomolecular 
condensates~\cite{Li2012,Banani2016,Harmon2017},
network concepts such as percolation 
transition~\cite{Harmon2017,Harmon2018,lassi,ipsen2020}
and graph theory~\cite{SandersCliff2020,Schmit2020} 
have been applied to provide rationalizations and theoretical formulations
of LLPS and gelation~\cite{LinBiochem2018}. In this perspective,
the nodes or ``stickers'' are folded or labile IDR interacting domains
and the LLPS or gelation transition is determined in large measure
by the number of connections (valence of attractive interactions) that 
can be effectuated between the nodes. Recent applications of this approach
include elucidating the role of RNA-binding valence in creating 
sufficient connectivity of a ribonucleoprotein network~\cite{SandersCliff2020}
in the assembly of stress granules~\cite{StressTaylor,StressAlberti}
and the competition between gelation and LLPS of the
cancer-related protein SPOP with its substrate~\cite{Schmit2020}.
\\

\noindent
{\bf Postsynaptic densities}

With these general considerations in mind, the present work aims to gain 
insight into the LLPS-driving interactions of 
two modular proteins, synaptic Ras GTPase-activating protein (SynGAP) and 
postsynaptic density protein 95 (PSD-95), which are major constituents of 
postsynaptic densities (PSDs)~\cite{Zeng2016}. Recent advances indicate that
biomolecular condensates play critical roles in neural function, probably
because of their ability to respond rapidly to stimuli, thus misregulated 
LLPS can lead to neurological 
diseases~\cite{takumi2018,fawzi2019,mingjie2020,mingjie2020b}. 
In particular, phase separation is important for proper communication between 
neurons as LLPS is involved in the assembly of synaptic vesicles (SVs) 
attached on presynaptic plasma 
membranes~\cite{Milovanovic2017, Milovanovic2018} 
as well as the PSDs beneath postsynaptic 
membranes~\cite{Zeng2016, Feng2018, Zeng2018a, Zeng2018b}.

Each PSD is a disc-shaped membraneless protein-rich compartment, an assembly 
made up of nucleic acids and thousands of different types of 
proteins~\cite{Zhuetal2016,takumi2018}. These include RNA binding 
proteins~\cite{jordan2012}, transmembrane layer proteins such as 
N-methyl D-aspartate receptor (NMDAR)~\cite{nmdar_ref}, 
actins in the cytoskeleton layer, 
scaffold proteins such as PDZ-domain-containing PSD-95s, which are 
membrane-associated guanylate kinases (MAGUKs)~\cite{maguk_ref}, 
guanylate-kinase (GK) associated protein (GKAP)~\cite{gkap_ref},
SH3 and multiple ankyrin repeat domain proteins (Shanks)~\cite{shank_ref},
and Homer family of adaptor proteins~\cite{homer_ref}, as well as
the SynGAP protein known to be highly enriched in 
PSDs~\cite{SynGAP_ref,Zeng2018a}.
Situated adjacent to postsynaptic membranes, PSDs can
exchange materials with the cytoplasm in synaptic spines and, accordingly, 
serves to provide spatial and temporal organization of the 
neurotransmitter receptors at the synapse~\cite{takumi2018}.
The functional importance of PSDs is highlighted by intriguing recent 
observations that PSDs are downsized during 
sleep~\cite{deVivo_etal2017,diering_etal2017}, which suggest that
synaptic strength is renormalized during sleep; and that
a SynGAP-PSD-95 condensate model of PSD~\cite{Zeng2016} can be disassembled 
by moderate hydrostatic pressure~\cite{RolandMingjie,RolandRev}, which points
to a possible biophysical origin of the high pressure neurological syndrome 
experienced by divers~\cite{talpalar2007}.

A minimal experimental model of PSD constructed using SynGAP and PSD-95---both
of which are very abundant in PSDs---has been observed to self-organize
into highly condensed, PSD-like droplets {\it in vitro}, offering the
first indication that LLPS is a significant biophysical underpinning of
PSD assembly~\cite{Zeng2016}.
Subsequently, more realistic {\it in vitro} models of PSD, encompassing
SynGAP and three scaffold proteins GKAP, Shank3, and Homer3 (in 
addition to PSD-95) as well as NR2B as another glutamate receptor,
have also been constructed as a versatile research platform~\cite{Zeng2018a}.
Here, as a first step toward elucidating the statistical mechanics of
PSD assembly, we focus on the simpler two-component SynGAP--PSD-95 construct, 
the same system we have utilized recently for studying the effect of
hydrostatic pressure on PSD stability~\cite{RolandMingjie}. SynGAP and PSD-95 
undergo LLPS together when mixed but each of the individual components, 
SynGAP or PSD-95 (up to 100 $\mu$M concentration each), does not phase 
separate by itself~\cite{Zeng2016}. It follows that the SynGAP--PSD-95 
condensate is a complex coacervate, the formation of which must involve 
favorable interactions between SynGAP and PSD-95 molecules.
\\

\noindent
{\bf LLPS of stoichiometric SynGAP/PSD-95 complexes
as a possible mechanism of PSD assembly}

Wildtype SynGAP has a long coiled-coil (CC) domain that can dock onto two 
other such domains in an intertwining manner resulting in a highly stable
SynGAP trimer, which we term $\Syn_3$ hereafter. The SynGAP trimer can, in 
turn, form a complex with two PSD-95 molecues~\cite{Zeng2016}. We refer
to this SynGAP--PSD-95 complex with a 3:2 stoichiometry, which is observed 
in dilute solution of SynGAP and PSD-95, as $\Syn_3\PSD_2$. Experimental 
studies indicate that formation of the trimer $\Syn_3$ is necessary for 
PSD-like complex coacervation, because a SynGAP mutant that abolishes 
SynGAP's ability to trimerize does not undergo LLPS in the presence of 
PSD-95~\cite{Zeng2016}. Taken together, these observations raise the 
tantalizing possibility that units of the $\Syn_3\PSD_2$ complex act as
the only nodes of favorable interactions in a SynGAP/PSD-95 condensate.
In other words, the condensate is stabilized solely by interactions among 
the $\Syn_3\PSD_2$ complexes.
Examples of how inter-$\Syn_3\PSD_2$ favorable interactions may
arise from known interactions among SynGAP and PSD-95 domains that stabilize 
individual $\Syn_3\PSD_2$ complexes \cite{Zeng2016,Zeng2018b} are 
provided in Fig.~1.
Recent experiments indicate that
the GK domain of a $\PSD$ molecule interacts favorably with  
one of the PDZ domains of another $\PSD$ molecule when that PDZ domain 
is bound to the one of the PDZ binding domains (PBMs) of a 
$\Syn_3$~\cite{Zeng2018b}. These GK--PDZ-PBM interactions can lead
to a network of favorably interacting $\Syn_3\PSD_2$ (Fig.~1b and c).
While such interactions may lead merely to a linear array of favorably 
interacting $\Syn_3\PSD_2$s (Fig.~1b) as previously 
envisioned~\cite{Zeng2018b}, they may also
result in a three-dimensional network of favorably interacting
$\Syn_3\PSD_2$s (Fig.~1c) and may therefore
stabilize a condensed phase of $\Syn_3\PSD_2$ complexes.
The hypothetical scenario 
that SynGAP/PSD-95 condensate is
stabilized entirely by favorable interactions among
units of $\Syn_3\PSD_2$ complexes represents a strong coupling between the 
$\Syn_3+2\PSD\rightleftarrows \Syn_3\PSD_2$ binding equilibrium
and LLPS-driving interactions. As such, it is instructive to ascertain,
theoretically, the implications of this assumed scenario on the predicted
LLPS properties. This knowledge would be useful, in general, not only for 
studying the SynGAP/PSD-95 systems but also other coacervate systems that 
involve a similar binding equilibrium, or a chemical equilibrium of reversible
reactions that takes an equivalent mathematical form as a binding equilibrium.
The theoretically predicted mesoscopic phase properties, such 
as co-existence curves and phase diagrams, may then be compared against 
corresponding experimental observations to assess
whether the assumed microscopic stoichiometric pattern of network 
interactions is indeed the case or, alternatively, auxiliary nodes of favorable 
interactions in addition to $\Syn_3\PSD_2$s are involved in stabilizing 
the condensed coacervate phase. 
\\

%%%%%%%

\noindent
{\bf Probing interaction networks in SynGAP/PSD-95 condensates
by theory and LLPS experiment}

To this end, we endeavor to develop a theoretical formulation for 
protein complex coacervation with an emphasis on the interactions among 
the folded protein domains under constraints similar to those imposed by 
the modular organization, multivalency and stoichiometry of the SynGAP/PSD-95 
system. In contrast, most of the recent attention of 
theoretical/computational developments on biomolecular condensates 
have been on IDPs and IDRs. While insights have been
gained into multiple-component LLPS by several recent theoretical/modeling
studies (e.g., Refs.~\cite{Feric2016,frenkel2017,Lin2017c,Harmon2018,LinBiochem2018,Paletal2021}), 
and some folded domains such as the helicase domain in LAF-1 have
been considered in explicit-chain simulations~\cite{DignonPLoS},
the preponderance of the efforts thus far are
on systems with a single IDP/IDR species.
For instance, sequence-dependent IDP LLPS has been studied using
analytical theory~\cite{Lin16,Kings}, coarse-grained explicit-chain
lattice~\cite{lassi,Das2018a,stefan2019}
and continuum~\cite{DignonPLoS,SumanPNAS2020,anders2020,JeetainPNAS2020}
as well as field-theoretic~\cite{McCarty2019,Paletal2021} simulations.
These efforts have focused on how LLPS propensity of heteropolymeric chain
molecules depends on the sequence of charged 
monomers~\cite{Lin16,Das2018b,Jonas2021} in general (as characterized, e.g., 
by quantitative charge pattern parameters~\cite{Das2013,Sawle2015}), 
effects of sequence patterns of hydrophobic~\cite{panag2020}, 
aromatic~\cite{RohitSci2020}, and other amino acid 
residues~\cite{DignonPLoS,SumanPNAS2020} on IDP phase behaviors, and 
the impact of sequence patterns on the phase properties of synthetic 
polymers~\cite{PerrySing2020,PerrySingRev2020}. 

The microscopic complexity of the SynGAP/PSD-95 system makes modeling
its LLPS in atomic details with quantitative precision infeasible. 
In order to gain insight into the essential biophysics of the system, we 
construct a highly simplified model that treats each $\Syn_3$ or $\PSD$ 
molecule as a single particle that can engage in certain interactions 
suggested by experimental observations. The model is for LLPS driven largely 
by folded domain association. As such, it is substantially different from 
the aforementioned theories for IDP LLPS that take into account the 
conformational diversity of IDP chains. In 
this respect, the conceptual underpinnings of the present model are akin to 
those of patchy particle models of 
colloids~\cite{wertheim1986,patchycolloid2017,langmuir2020}, which
have been applied recently to study phase behaviors of
folded proteins such as lens crystallins~\cite{vojko2016},
ribonucleoprotein droplets in which both the proteins and RNA
are modeled as patchy articles~\cite{hxzhou2018}, 
as well as
the relationship between patchy-particle and coarse-grained
explicit-chain models of biomolecular LLPSs including assembly of
multicomponent condensates~\cite{RosanaCG-PNAS2020}.
The model developed herein is structurally and energetically more simplified 
than patchy particle formulations in that the valencies of 
$\Syn_3$ and $\PSD$ are now treated in a mean-field manner 
without considering the interactions' anisotropic directionality. 
This simplicity notwithstanding, our tractable model provides 
a critical assessment of SynGAP/PSD-95 coacervation scenarios 
because the model features a possible coupling between the formation of the 
3:2 $\Syn_3\PSD_2$ complex and phase separation. We find that
the constraints imposed by an exclusively complex-driven phase separation entail
drastically different $\Syn_3$ and $\PSD$ compositions of coexisting phases
from those predicted by classical Flory-Huggins 
theories~\cite{flory1953,deGennes1979} 
in the absence of such coupling. Therefore, whether SynGAP/PSD-95 coacervation
is driven solely by interactions among $\Syn_3\PSD_2$ complexes can be
assessed by comparing our theoretical predictions against experiment.
With this recognition, we have now conducted extensive 
experimental measurements of $\Syn_3$/$\PSD$ compositions of coexisting 
phases. The results are consistent with classical Flory-Huggins theory 
but not the exclusively complex-driven LLPS theory, indicating a
hitherto unknown prevalence of auxiliary interactions beside those among 
the $\Syn_3\PSD_2$ complexes in the SynGAP/PSD-95 
condensed phase. Details of these findings and their ramifications are 
provided below.
\\

%%%%%%%%%%%%%%%%%%%%%%%%%%%%%%%%%%%%%%%%%%%%%%%%%%%%%%%%%%%%%%%%%%%%%%%%%%%%
%%%%%%%%%%%%%%%%%%%%%%%%%%%%%%%%%%%%%%%%%%%%%%%%%%%%%%%%%%%%%%%%%%%%%%%%%%%%%%

\noindent
{\bf MODELS, MATERIALS, AND METHODS}
\\

\noindent
{\bf A general mean-field theoretical framework for
stoichiometric complex-driven LLPS}

Before delving into the particulars of the SynGAP/PSD-95 system, 
it is useful and instructive to first construct a general LLPS theory for 
a solution containing two species of solutes $A$ and $B$ that are
capable of forming a stoichiometric complex $A_xB_y$ (where $x$, $y$ are
two fixed positive integers) and whose complex coacervation is 
contingent upon, i.e., coupled to, the formation of such complexes.
The formation and dissociation of the complex may be expressed as a
reversible binding or chemical reaction:
\begin{equation}
x A + y B \rightleftarrows A_x B_y
\; .
	\label{eq:AxBy}
\end{equation}
Now let the total solution volume be $\Vol$, and the number of solutes
$A$ and $B$ in the solution be, respectively, $n_A$ and $n_B$. Proceeding 
with a Flory-Huggins (FH)-type lattice argument \cite{flory1953,chandill1994},
we assume for simplicity that individual $A$ and $B$ solute molecules
occupy equal volume, denoted by $v_{\rm s}$, and discretize the solution volume
conceptually to a total of $M=\Vol/v_{\rm s}$ lattice sites. 
Accordingly, a lattice 
site in the system can be occupied by an $A$ or $B$ solute molecule it its
entirety, part of an $A_x B_y$ complex, or an entire solvent molecule. In 
other words, each $A$, $B$, or solvent molecule takes up one lattice site, 
whereas each $A_xB_y$ complex occupies a contiguous chunk of $x+y$ sites.

Consider a situation in which $m$ complexes are formed in the solution.
The binding reaction and the configurational freedom of such a state is 
described by the partition function
\begin{equation}
Z_{\rm bind}(m) = \frac{M! e^{m\eng/\kB T}}{m!(n_A-xm)!(n_B-ym)! (M-n_A-n_B)!}
\; ,
\label{eq:Zbind}
\end{equation}
where $-\eng < 0$ is the free energy associated with the favorable 
complex-forming binding (forward) reaction in Eq.~\ref{eq:AxBy}, $\kB$ 
is Boltzmann constant, and $T$ is absolute temperature. The free energy 
$\eng$ does not account for the part of translational entropy described
by the combinatoric factors in Eq.~\ref{eq:Zbind}. Nonetheless, in general 
$\eng$ can contain enthalpic as well as entropic components with entropic 
contributions from, e.g., change in solvent orientational entropy 
associated with the binding reaction.
It should also be noted that
for notational simplicity a standard FH factor, $[(z-1)/M]^{m(x+y-1)}$, 
to account for the spatial contiguity of each of the $A_x B_y$ complex
(where $z$ is coordination number of the lattice for a
flexible, polymer-like complex and $z-1\rightarrow 1$ 
for a rigid complex)~\cite{chandill1994}) is omitted in Eq.~\ref{eq:Zbind}
because the factor may be formally absorbed into the entropic component of 
$\eng$ by redefining $\eng \rightarrow \eng + \kB T (x+y-1)\ln [(z-1)/M]$.

We further assume that LLPS of our model system is driven
by short spatial range contact-like interactions between individual
bound solute $A$ or $B$ molecules from different $A_xB_y$ complexes.
For each $m$-complex state, the energetic contribution for these
interactions is given by the following free energy 
\begin{equation}
-\kB T \ln Z_{\rm LLPS}(m) 
= - \frac{z\engc}{M}\left( \begin{array}{c} m \\ 2 \end{array} \right)(x+y)^2 
+ O(m^3)
\; ,
\label{eq:Zllps}
\end{equation}
where $Z_{\rm LLPS}(m)$ is the pertinent partition function and $\engc$
is the contact free energy of an $A$--$A$, $B$--$B$, or $A$--$B$ interaction
(which are assuned to carry the same $\engc$ for simplicity). In situations
where these interactions are relatively weak, as is apparent the case
for the SynGAP/PSD-95 system according to a recent 
analysis of experimental data on the system's sensitivity to
hydrostatic pressure \cite{RolandMingjie}, one may consider only the
pairwise $O(m^2)$ contribution in Eq.~\ref{eq:Zllps}, which corresponds
to the lowest-order interaction term in coordinate-space polymer lattice 
cluster theory~\cite{Baker1993}. 

It should be emphasized that the $\engc$ for $A$--$A$, $B$--$B$, and $A$--$B$
contacts in the present formulation is only for inter-complex interactions
between different $A_xB_y$ units. The parameter $\engc$ is not involved
in the assembly of an individual $A_xB_y$ complex, 
which is stipulated separately by a chemical equilibrium [Eq.~\ref{eq:AxBy}]
as will be described further below. In the present mean-field formulation,
each unit of $A_xB_y$ interacts as a whole in an isotropic manner. 
In this context, assigning different energies to inter-complex $A$--$A$, 
$B$--$B$, and $A$--$B$ contacts would only amount, mathematically, to a 
redefinition of $\engc$, which is effectively a weighted average 
inter-complex energies over different types of contacts between the complexes' 
$A$ and $B$ constituents. Going forward, it would be interesting in future 
investigations to extend the present approach to stucturally and energetically
more realistic models that allow for anisotropic interactions, such as 
patchy particle models~\cite{RosanaCG-PNAS2020}.  Under those extended
frameworks, it would then be useful to study the effects of different
inter-complex energies for $A$--$A$, $B$--$B$, and $A$--$B$ contacts.

Here, with $Z_{\rm bind}$ and $Z_{\rm LLPS}$ in place, the total partition
function $\Zall$ for our model solution system that takes into account all
possible $m$-complex states is given by
\begin{equation}
\Zall = \sum_{m=0}^{m_{\rm max}}Z_{\rm bind}(m)Z_{\rm LLPS}(m) 
\equiv \sum_{m=0}^{m_{\rm max}} e^{-M f(m)}
\; ,
\label{eq:Zall}
\end{equation}
where $m_{\rm max} = \lfloor{\rm min}\{n_A/x, n_B/y\}\rfloor$ is
the maximum number of $A_xB_y$ complexes possible for a given solution system
and $f(m)$ is the $m$-state free energy per lattice site in units of
$\kB T$. Application of Stirling's approximation for the factorials
to Eqs.~\ref{eq:Zbind} and \ref{eq:Zllps} leads to the following 
approximate expression for $Mf(m)$:
\begin{equation}
\begin{aligned}
& M f(m)  \\
& \approx
(n_A-x m)\ln(n_A-x m) + (n_B-y m)\ln(n_B-y m) \\
& \quad +  (M-n_A-n_B)\ln(M-n_A-n_B) \\
& \quad + m \ln m + m\left(  x + y -1 -\frac{\eng}{\kB T}\right) \\
& \quad - \chi \frac{[m(x+y)]^2}{M} - M\ln M
\; ,
\end{aligned}
	\label{eq:f_m}
\end{equation}
where $\chi = z\engc/(2\kB T)$ is an effective FH 
parameter~\cite{Baker1993, Bawendi1987}. In arriving at Eq.~\ref{eq:f_m},
we have neglected $O(1/M)$ and $O([\ln M]/M)$ contributions to $f(m)$ 
because these terms vanish in the $M\rightarrow\infty$ thermodynamic limit.
Accordingly, the $\chi m(x+y)^2$ term from Eq.~\ref{eq:Zllps}
and terms from the $(2\pi n)^{1/2}$ part of the
Stirling approximation $n!\approx (2\pi n)^{1/2}n^n e^{-n}$
are not included in Eq.~\ref{eq:f_m}.

Following common notation in FH phase separation theory, 
we define $\phi_A\equiv n_A/M$ and $\phi_B\equiv n_B/M$ as overall
volume fractions for $A$ and $B$, respectively,
$\psi \equiv m(x+y)/M$ as volume fraction for $A_x B_y$, and
$\psi_A \equiv \phi_A - x\psi/(x+y)$ and $\psi_B \equiv \phi_B - y\psi/(x+y)$
as volume fractions, respectively, of unbound $A$ and $B$
(i.e., which do not form $A_x B_y$). Using this notation, Eq.~\ref{eq:f_m} 
may be rewritten as
\begin{equation}
\begin{aligned}
f(\psi) 
%= & \left( \phi_A \!-\! \frac{x\psi}{x+y}\right)\ln \left( \phi_A \!-\! \frac{x\psi}{x+y}\right) \\
%& + \left( \phi_B \!-\! \frac{y\psi}{x+y}\right)\ln\left( \phi_B \!-\! \frac{y\psi}{x+y}\right) \\
= & \psi_A \ln \psi_A + \psi_B \ln \psi_B + \frac{\psi}{x+y}\ln\frac{\psi}{x+y}\\
& + (1-\phi_A-\phi_B)\ln(1-\phi_A-\phi_B) \\
& + \frac{\psi}{x+y} \left(  \ln\Kdr^0 + x + y -1 \right)   - \chi \psi^2 
\; ,
\\
\label{eq:fpsi}
\end{aligned}
\end{equation}
where 
\begin{equation}
\Kdr^0 \equiv \frac{e^{-\eng/\kB T}}{M^{x+y-1}}
\end{equation}
is a reduced (dimensionless) dissociation constant of $A_x B_y$ in 
dilute solution (the ``0'' superscript symbolizes dilute solution), the unit 
of which will be clarified below.

We are now ready to derive an approximate expression for the system
described by $\Zall$ by replacing the summation in Eq.~\ref{eq:Zall}
with a single term at the saddle point $\psib$ of $f(\psi)$. In other words,
the free energy of the system
\begin{equation}
\Fall = -\kB T \ln \Zall \approx M f(\psib) \; ,
\end{equation}
where $\psib$ is obtained by solving
\begin{equation}
\begin{aligned}
0 = \left.\frac{\partial f(\psi) }{\partial \psi}
\right|_{\psi=\psib} = & 
-\frac{x}{x+y} \ln\psib_A -\frac{y}{x+y} \ln\psib_B   \\
& + \frac{1}{x+y}\left[ \ln \frac{\psib}{x+y} + \ln\Kdr^0 \right] - 2\chi\psib
\; . \\
	\label{eq:dfdpsi=0} 
\end{aligned}
\end{equation}
Multiplying every term on the right hand side of Eq.~\ref{eq:dfdpsi=0} 
by $x+y$ and taking exponential of all terms result in the equivalent condition
\begin{equation}
\frac{\psib_A^{\;x} \psib_B^{\;y} }{[\psib/(x+y)]} = \Kdr^0 
e^{-2\chi(x+y)\psib}
\; ,
	\label{eq:AB_equilibrium}
\end{equation}
which can be further rewritten as
\begin{equation}
\frac{[A]^x[B]^y}{[A_xB_y]} = K_{\rm d}^0 e^{-2\chi v_{\rm s}(x+y)^2[A_xB_y]  }
	\label{eq:AB_equilibrium}
\end{equation}
where $[A]$, $[B]$, $[A_xB_y]$ are the equilibrium concentrations,
respectively, of $A$, $B$, $A_xB_y$ and
$K_{\rm d}^0$ is the dissociation constant of $A_xB_y$ in dilute
solution. These quantities are given by
\begin{subequations}
\begin{align}
[A_xB_y] & = \frac{\psib}{v_{\rm s}(x+y)} \equiv \frac{\overline{m}}{\Vol} \\
[A] & = \frac {\overline{\psi}_A}{v} 
= [A_{\rm T}] - x [A_xB_y] = \frac{n_A}{\Vol}-x \frac{\overline{m}}{\Vol} \\
[B] & = \frac {\overline{\psi}_B}{v}=
[B_{\rm T}] - y [A_xB_y] = \frac{n_B}{\Vol} -y \frac{\overline{m}}{\Vol} \\
K_{\rm d}^0 & = \Kdr^0 v_{\rm s}^{x+y-1} = \frac {e^{-\eng/\kB T}}{V^{x+y-1}}
\; ,
\end{align}
\end{subequations}
where $[A_{\rm T}]$ and $[B_{\rm T}]$ are total concentrations of $A$
and $B$, irrespective of 
whether they are unbound or part of an $A_xB_y$ complex.
As expected, when $\chi=0$, Eq.~\ref{eq:AB_equilibrium} reduces
to the equilibrium equation
for the binding reaction in Eq.~\ref{eq:AxBy} for dilute solution.
In contrast, when $\chi\neq 0$, i.e., in the presence of the LLPS-driving
interactions, the dissociation constant of $A_xB_y$ defined
by $[A]^x[B]^y/[A_xB_y]$ in Eq.~\ref{eq:AB_equilibrium}
decreases exponentially as $\chi$ increases. This is because
the individual $A$ and $B$ solutes in an $A_xB_y$ complex cannot 
dissociate from one another as long as the $A_x B_y$ complex 
is interacting with another $A_xB_y$ complex via the LLPS-driving 
FH interaction. This constraint results in an enhanced binding affinity between 
$A$ and $B$ relative to that in a dilute solution where
the LLPS-driving inter-complex interactions are absent.
\\

\noindent{\bf Experimental materials and methods}

{\it Protein expression and purification.} 
The PSD-95 construct containing PDZ, SH3
and GK tandem (aa R306-L721 of the protein, referred to as PSD-95 in this
paper) was PCR amplified from human cDNA library 
(NCBI: NP${\underline{\phantom{o}}}$001122299). The
SynGAP construct containing a trimeric coiled-coil (CC) domain and a PSD-95 PSG
binding domain (PBM) (encoding aa A1147-V1308, lacking 1192V-1193K and
1293E-1295G of the protein, referred to as SynGAP in this paper) was PCR
amplified from mouse cDNA library (UniProt: J3QQ18)~\cite{Zeng2016}. The
genes encoding PSD-95 and SynGAP was individually cloned into vectors
containing N-terminal His6-affinity tag. Recombinant proteins were expressed in
{\it Escherichia coli} 
BL21(DE3) cells in LB medium at 16$^\circ$C overnight and purified
using a nickel-NTA agarose affinity column followed by size-exclusion
chromatography (Superdex 200) with a column buffer containing 50mM Tris, pH8.2,
100mM NaCl, 1mM EDTA, 1mM DTT. The fusion tag of each protein was cleaved by
HRV 3C protease and separated by another step of size-exclusion chromatography. 

{\it Sedimentation-based assay of phase separation and phase diagram.} 
All proteins were pre-cleared through high-speed centrifugation 
(16873$\times$g 10 min)
at 25$^\circ$C before sedimentation assay. Proteins were mixed at each designed
concentration in 50 $\mu$l total volume. After 10-min equilibration at room
temperature, the mixture was centrifuged at 16873$\times$g at 
25$^\circ$C for 10 min. The
supernatant was collected, and pellet was thoroughly resuspended in 50 $\mu$l
buffer. Proteins from supernatant and pellet fractions were analyzed through
SDS-PAGE with Coomassie blue R250 staining.

The phase diagram was constructed from the sedimentation-based phase separation
assay. Phase separation of a SynGAP/PSD-95 mixture was considered to occur if
the concentrations of SynGAP and PSD-95 recovered from the pellet of the
mixture were higher than the pellet fraction of the individual protein alone in
each sedimentation assay.

{\it Quantification of protein concentration in the condensed phase.}
Concentrations of PSD-95 in the condensed phase were measured according to the
previous reported protocol~\cite{Zeng2018a}. Briefly, Cy3-labeled PSD-95 was
diluted into the final concentration of 1\% by mixing with unlabeled protein. 
1\% Cy3-labeled PSD-95 was further mixed with SynGAP to form condensates. The
mixture was injected into a 96-well glass-bottomed plate (Thermo Fisher).
Confocal fluorescence images were captured using Zeiss LSM880 confocal
microscope with a 63$\times$ objective lens. 
The confocal slice images spanning the
middle of each droplet were used to quantify the concentration of PSD-95. 

To calculate concentration of PSD-95 in each droplet, a standard calibration
curve was generated. Cy3-labeled PSD-95 alone in dilute solutions at different
concentrations were added into a glass-bottomed plate. The fluorescence
intensity of each confocal slice image was measured using the same settings as
that for the droplet quantifications. According to the standard curve,
fluorescence intensity of PSD-95 in each condensate can be converted into
absolute molar concentration.
\\

\noindent
{\bf RESULTS AND DISCUSSION}
\\

\noindent
{\bf A mean-field theory of stoichiometric complex-driven SynGAP/PSD-95 LLPS}

We are now in a position to apply the general theory developed above to the
SynGAP/PSD-95 system, with the amino acid sequence information of the SynGAP 
and PSD-95 constructs (157 and 416 residues, respectively) specified 
above under the ``Experimental materials and 
method'' heading. For wildtype SynGAP, because of the extremely strong 
propensity to trimerize \cite{Zeng2016}, we consider, as a very good
approximation, that all SynGAP molecules exist as part of $\Syn_3$ trimers 
in solution. Accordingly, we set $A=\Syn_3$, $B=\PSD$, $x=1$, and $y=2$ 
(Fig.~2a) in the formalism of the last section to obtain from
Eq.~\ref{eq:fpsi} the free energy for the SynGAP/PSD-95 system:
\begin{equation}
\begin{aligned}
f(\phi_{\Syn_3}, \phi_\PSD) 
= & \psib_{\Syn_3}\ln\psib_{\Syn_3} +  \psi_{\PSD}\ln\psib_{\PSD}  
+ \frac{\psib}{3}\ln\frac{\psib}{3} \\
& + (1-\phi_{\Syn_3}-\phi_\PSD)\ln (1-\phi_{\Syn_3}-\phi_\PSD)  \\
& + \frac{\psib}{3}\left[ \ln\Kdr^0 + 2 \right] - \chi \psib^2
\; ,
\end{aligned}
	\label{eq:final}
\end{equation}
where 
\begin{subequations}
\begin{align}
\psib_{\Syn_3} = & \phi_{\Syn_3}- \psib/3, \\
\psib_{\PSD} = &  \phi_{\PSD_3}- 2\psib/3, \\
\frac{\psib}{3}\Kdr^0  = & \psib_{\Syn_3}\psib_{\PSD}^2 e^{6\psib} 
\; 
\label{eq:2step_Kd}
\end{align}
\end{subequations}
follow from Eq.~\ref{eq:AB_equilibrium}.
The LLPS-driving interaction network envisioned by this formulation
is schematically illustrated in Fig.~2b.

The corresponding phase-coexistence behaviors are solved by the numerical 
method described in Ref.~\cite{Lin2017c}. In general, for a solution with
overall average protein volume fractions $(\phi_{\Syn_3}^0, \phi_{\PSD}^0)$, the
free energy without phase separation is given by 
\begin{equation}
f_{\rm bulk} = f(\phi_{\Syn_3}^0, \phi_{\PSD}^0)
\; .
\end{equation}
Whether the system phase separates is determined by the difference 
between $f_{\rm bulk}$ and the free energy $f_{\rm sep}$ of a phase-separated 
system with two coexisting phases labeled by $\alpha$ and $\beta$, viz.,
\begin{equation}
f_{\rm sep} = v f(\phi_{\Syn_3}^\alpha, \phi_{\PSD}^\alpha) 
			+ (1-v) f(\phi_{\Syn_3}^\beta, \phi_{\PSD}^\beta)
\; ,
\label{eq:veq1}
\end{equation}
where $v$ is the fraction of the total system volume in phase $\alpha$.
By definition, $v$ satisfies $0< v < 1$ and the volume conservation conditions
\begin{subequations}
\begin{align}
v \phi_{\Syn_3}^\alpha + (1-v) \phi_{\Syn_3}^\beta & = \phi_{\Syn_3}^0 \; ,\\
v  \phi_{\PSD}^\alpha + (1-v) \phi_{\PSD}^\beta & = \phi_{\PSD}^0 \; ,
\end{align}
\end{subequations}
which, in turn, can be used to 
relate volume fractions in the $\beta$ phase to those in the $\alpha$ phase,
allowing $f_{\rm sep}$ for overall volume fractions
$\phi_{\Syn_3}^0$, $\phi_{\PSD}^0$ to be rewritten as a function of $v$, 
$\phi_{\Syn_3}^\alpha$, and $\phi_{\PSD}^\alpha$:
\begin{equation}
\begin{aligned}
\!\!\!
f_{\rm sep} (v, \phi_{\Syn_3}^\alpha, \phi_{\PSD}^\alpha)
= & v f(\phi_{\Syn_3}^\alpha, \phi_{\PSD}^\alpha) \\
   & + (1-v) f\left(
   		\frac{\phi_{\Syn_3}^0\!-\!v\phi_{\Syn_3}^\alpha}{1-v},
             	\frac{\phi_{\PSD}^0\!-\!v\phi_{\PSD}^\alpha}{1-v}
             \right)
\; .
\end{aligned}
\label{eq:veq3}
\end{equation}
The system will phase separate if there exists at least a set of $(v,
\phi_{\Syn_3}^\alpha, \phi_{\PSD}^\alpha)$ values for which 
$ f_{\rm sep} < f_{\rm bulk}$. In that case, the final protein volume 
fractions---i.e. the binary coexistence phase boundary---are the volume 
fractions that minimize $f_{\rm sep}$. This minimization is achieved
numerically by implementing a three-variable sequential least squares 
programming (SLSQP) algorithm~\cite{Kraft1988} using 
the {\tt scipy.optimize.minimize} function in
Scipy, a  Python-based numerical package for scientific
computation~\cite{Scipy}. To ensure that there is no numerical errors
in our calculation, we check the optimized SLSQP solutions 
against the following equations for chemical potential 
balance of coexisting phases to confirm that these conditions
are indeed satisfied:
\begin{subequations}
\begin{align}
f_{\Syn_3}'^{\alpha} = & f_{\Syn_3}'^{\beta} \; , \\ 
f_{\PSD}'^{\alpha} = & f_{\PSD}'^{\beta} \; ,   \\
\mu_{\rm w}^\alpha = & \mu_{\rm w}^\beta \; ,
\end{align}
\end{subequations}
where 
\begin{subequations}
\begin{align}
f^y \equiv & f(\phi_{\Syn_3}^y, \phi_{\PSD}^y), \\
f'^y_x \equiv 
&\left. \frac{\partial f(\phi_{\Syn_3}, \phi_{\PSD})}{\partial \phi_x} 
		\right|_{(\phi_{\Syn_3}, 
\phi_{\PSD})=(\phi_{\Syn_3}^y, \phi_{\PSD}^y)} \; , \\
\mu_{\rm w}^y = & f^y - \phi_{\Syn_3}^y f'^y_{\Syn_3} 
- \phi_{\PSD}^y f'^y_{\PSD},
\end{align}
\end{subequations}
with $x \in \{\Syn_3, \PSD\}$ and $y \in \{ \alpha, \beta \}$.

The trend of phase behaviors predicted by the free energy given by
Eq.~\ref{eq:final} for this model of stoichiometric complex-driven LLPS 
is illustrated by the phase diagram in Fig.~2c, which features an
L-shape phase-separated regime (bound by the light blue solid lines
and the diagonal black solid line with slope $=-1$). Apparently,
a hallmark of stoichiometric complex-driven LLPS is a trend of
converging tie lines (dark blue dashed lines in Fig.~2c) from the 
dilute, protein-depleted phase boundary (large light-blue ``L'' in Fig.~2c) to 
the condensed, protein-rich phase boundary (small light-blue ``L'').
When the overall concentrations (volume fractions) of
$\Syn_3$ and $\PSD$ are in the dilute-solution stoichiometric ratio of 1:2 
($\phi_{\Syn_3}^0/\phi_\PSD^0=1/2$), the tie line has slope $=1/2$, 
coinciding exactly with the black solid line in Fig.~2c with the same slope,
indicating that this 1:2 stoichiometric ratio is maintained in both the 
dilute and condensed phases. When $\phi_{\Syn_3}^0/\phi_\PSD^0\neq 1/2$,
the molecular species ($\Syn_3$ or $\PSD$) that is in excess of the $1/2$ ratio
has no effect on phase separation because they do not generate more
$\Syn_3\PSD_2$ complexes and, by model construction, only interactions 
among $\Syn_3\PSD_2$ complexes are capable of driving LLPS.
Consequently, the tie lines connecting coexisting dilute and condensed 
phases are converging toward the $\phi_{\Syn_3}/\phi_\PSD=1/2$ line, meaning
that the condensed phase is always closer than the dilute phase to the
$\phi_{\Syn_3}/\phi_\PSD=1/2$ stoichiometric ratio if 
$\phi_{\Syn_3}^0/\phi_\PSD^0\neq 1/2$ because of the larger number of
favorable interactions among $\Syn_3\PSD_2$ complexes in the condensed phase.
With this ramification of a hypothetical stoichiometric complex-driven 
LLPS elucidated, we now proceed to test experimentally whether
the real SynGAP/PSD-95 system indeed undergoes such a complex 
coacervation process.
\\

\noindent
{\bf Experimental measurements of the PSD-95 concentrations
in the condensed phase of SynGAP/PSD-95 mixtures}

Our previous study showed that SynGAP and PSD-95 could undergo phase
separation together. In the dilute phase, the two proteins form a 
homogenous 3:2 complex~\cite{Zeng2016}.
However, the binding mode of SynGAP and PSD-95 in the
condensed phase is unknown. To explore whether the two proteins also form a
stoichiometric 3:2 complex in condensed phase, we measured concentrations of
the two proteins in dilute phase and in condensed phase (labeled in Fig.~3a as
``S'' for supernatant and ``P'' for pellet, respectively, as 
in Fig.~4 of ref.~\cite{Zeng2016}, not
to be confused with the notation $\Syn$ for SynGAP and $\PSD$ for PSD-95). 
SynGAP and PSD-95 were mixed at different molar ratios. Sedimentation-based 
assay was used to assess their distributions in the dilute and condensed 
phases by SDS-PAGE with Coomassie blue staining (Fig.~3a). 
The percentages of SynGAP and PSD-95 
recovered in the pellet were analyzed. A phase diagram of
the SynGAP/PSD-95 system was then constructed according to the data 
obtained from the sedimentation-based assay. Here, the phase diagram in 
Fig.~3b shows that complex coacervation of SynGAP and PSD-95 is highly
sensitive to the concentrations of the two proteins. When concentrations of
the two components were higher than certain threshold values, phase 
separation of the mixtures occurred persistently. 
Interestingly, when fixing the concentration of
one component at a value below its threshold value, increasing the
concentration of the other component could also induce phase separation. For
instance, there was no phase separation when SynGAP was
at 30 $\mu$M and PSD-95 was at 20 $\mu$M.
However, when the concentration of SynGAP was increased to 120 $\mu$M with
[PSD-95] unchanged at 20 $\mu$M, phase separation of the mixture was
observed (Fig.~3b). This observation suggests that SynGAP and PSD-95 interacted 
with each other not only through the PDZ domain of PSD-95 and
the PBM of SynGAP, but the interactions might 
also involve other sites of the two proteins. 
To quantitatively measure the absolute concentration of PSD-95 in the 
condensed phase so as to enable the theoretical analysis below that provides
a clearer answer to this basic question, we used a confocal fluorescence 
image-based method developed in our earlier study~\cite{Zeng2018a} (Fig.~3c,d). 
The fluorescence intensity of PSD-95 in
the condensed droplets under each condition was converted into absolute
concentration of the protein based on the standard curve of Cy3-labeled PSD-95
obtained in dilute solutions (Fig.~3c). The derived absolute concentrations
of PSD-95 under the conditions shown in the phase diagram of the PSD-95 and
SynGAP mixtures (Fig.~3b) are shown by the heat map in Fig.~3e.
\\

\noindent
{\bf Experimental concentrations of SynGAP and PSD-95 in
coexisting condensed and dilute phases}

The next step in our investigation is to analyze the
experimental data to determine whether they are consistent with 
the predicted behaviors in Fig.~2c for the hypothetical scenario
in which LLPS is driven solely by favorable interactions among 
stoichiometric $\Syn_3\PSD_2$ complexes. Consider an aqueous solution of 
$\Syn$ and $\PSD$ with initial concentrations $[\Syn^0]$ and $[\PSD^0]$. 
Using a notation similar to that in Eqs.~\ref{eq:veq1}--\ref{eq:veq3}
for this system when it has undergone phase separation, 
we define $v$ as the fraction of total volume in the dilute phase
and thus $1-v$ is the fraction of total volume in the condensed phase.
It follows from the conservation of $\Syn$ and $\PSD$ that
\begin{subequations}
\begin{align}
[\Syn^0] = & v[\Syn^{\rm dil}] + (1-v)[\Syn^{\rm cond}]  \label{eq:syn_ps} \\
[\PSD^0] = & v[\PSD^{\rm dil}] + (1-v)[\PSD^{\rm cond}] \label{eq:psg_ps} \;
\end{align}
\end{subequations}
where quantities in the dilute and condensed phases are labeled, 
respectively, by the superscripts ``dil'' and ``cond''. Note that we have 
defined the concentrations in Eqs.~\ref{eq:syn_ps} and \ref{eq:psg_ps}
purposely to correspond to those measured in the experiments (Fig.~3) in
that they refer to the total concentrations of $\Syn$ and $\PSD$ molecules
or their concentrations in the dilute and condensed phases irrespective 
of their binding status. Nonetheless, based on prior experiments~\cite{Zeng2016}
as noted in the introductory discussion above, we recognize that
virtually all $\Syn$ molecules are in the form of $\Syn_3$ under the present
experimental conditions such that $[\Syn^0]\approx 3[\Syn_3^0]$, 
$[\Syn^{\rm dil}]\approx 3[\Syn^{\rm dil}_3]$, and 
$[\Syn^{\rm cond}]\approx 3[\Syn^{\rm cond}_3]$ for the concentrations, 
whereas the volume fraction $\phi_\Syn\approx\phi_{\Syn_3}$. 

In order to construct a binary phase diagram, our goal here is to determine
$[\Syn^{\rm dil}]$, $[\Syn^{\rm cond}]$, $[\PSD^{\rm dil}]$, and 
$[\PSD^{\rm cond}]$ from two sets of experiments: (i) the 
121 centrifugation-based assays providing the fractional numbers 
of protein molecules in the condensed phase, denoted here as
\begin{subequations}
\begin{align}
\gamma_{\Syn} \equiv & \frac{(1-v)[\Syn^{\rm cond}] }{[\Syn^0]}
\; , \label{eq:gamma_syn}\\
\gamma_{\PSD} \equiv & \frac{(1-v)[\PSD^{\rm cond}] }{[\PSD^0]} 
\label{eq:gamma_psg} \; ,
\end{align}
\end{subequations}
and (ii) the 25 confocal-microscopy measurements affording
the condensed phase concentration, $[\PSD^{\rm cond}]$, of PSD-95 ($\PSD$), 
which may be identify, aside from a proportionality constant, to the 
volume fraction of $\PSD$:
\begin{equation}
\phi_{\PSD}^{\rm cond} \propto [\PSD^{\rm cond}] \; .
\end{equation}
Since the $\gamma_\Syn$ and $\gamma_\PSD$ in Eqs.~\ref{eq:gamma_syn} and 
\ref{eq:gamma_psg} determined from centriguation measurements (two experimental
data points per condition) are by themselves insufficient to determine both 
$[\Syn^{\rm dil}]$ and $[\PSD^{\rm dil}]$ as well as
$v$ (three variables) in Eqs.~\ref{eq:syn_ps} and \ref{eq:psg_ps},
we can only make the full determination of dilute- and condensed-phase
protein concentrations for the 25 conditions for which confocal microscopy
data are also available, with
\begin{align}
[\Syn^0] = & 30, 60, 90, 120, 180 \; \mu {\rm M} \; , \nonumber\\
[\PSD^0] = & 20,40,60,80,120 \; \mu {\rm M} \; . \nonumber
\end{align}
For each of these given conditions, we may first use Eq.~\ref{eq:gamma_psg} to
obtain 
\begin{equation}
v = 1 - \gamma_\PSD \frac{ [\PSD^0]}{[\PSD^{\rm cond}]}
\; 
\end{equation}
by using $\gamma_\PSD$ from centrifugation measurements and 
$[\PSD^{\rm cond}]$ from confocal microscopy. Once $v$ is determined,
$[\PSD^{\rm dil}]$ follows from Eq.~\ref{eq:psg_ps} as
\begin{equation} 
[\PSD^{\rm dil}] = \frac{[\PSD^0](1-\gamma_\PSD)}{v} \; ,
\end{equation}
and the SynGAP concentrations in the dilute and condensed phases 
follow, respectively, from 
Eqs.~\ref{eq:syn_ps} and~\ref{eq:gamma_syn} as 
\begin{subequations} 
\begin{align}
[\Syn^{\rm dil}] =  & \frac{[\Syn^0](1-\gamma_\Syn)}{v} \; , \\
[\Syn^{\rm cond}] = & \gamma_\Syn\frac {[\Syn^0]}{1-v} \; ,
\end{align}
\end{subequations} 
by using the solved value of $v$ and $\gamma_\Syn$ from 
centrifugation measurements.
\\

\noindent
{\bf General trend of binary SynGAP/PSD-95 phase behaviors 
inferred from experimental data is not consistent with the stoichiometric 
complex-driven LLPS scenario}

The above-described analysis of experimental data yields the
$([\PSD^{\rm dil}],[\Syn^{\rm dil}])$ and
$([\PSD^{\rm cond}],[\Syn^{\rm cond}])$ data points in Fig.~4 (orange
and blue dots) as well as
the tie lines connecting their coexisting values (dashed lines in Fig.~4).
Since the protein concentrations of the initial states in the experiments 
are very close to those of the dilute phases and the dilute-phase concentrations
are not measured directly, there is appreciable scatter in the
$([\PSD^{\rm dil}],[\Syn^{\rm dil}])$ and
$([\PSD^{\rm cond}],[\Syn^{\rm cond}])$ 
data points. 
This is because when initial concentrations of $\Syn$ 
and $\PSD$ are low, the droplets formed through phase
separation are very small. 
Thickness of individual droplet is often smaller than confocal optical
section thickness ($\sim 0.9 \mu$m). 
In such situations, the fluorescence intensity measured by confocal 
microscope is not only affected by fluorescence signal in the droplet but
also interfered by the glass of the coverslip and labeled proteins in solution. 
This basic limitation precludes an accurate quantification of the
uncertainties in measured condensed-phase concentrations when 
overall protein concentrations are low, making
it impractical to construct a smooth phase boundary from the plotted 
data. In future investigations, this difficulty can possibly be 
overcome by using
solutions that lead to larger droplets, in which case the quality of
the data may be evaluated by comparing signal intensities for different Z 
stacks obtained by confocal microscopy. For instance, if the thickness 
of droplet is larger than the confocal optical section thickness, the 
fluorescence intensities of several layers spanning the middle
of each droplet should be comparable (see, e.g., Fig.~3 of
ref.~\cite{Zeng2018a}).

Experimental limitations notwithstanding, one unmistakably
clear message from Fig.~4a is that the tie lines diverge from 
the protein-depleted phases (orange data points in Fig.~4a) 
to the protein-rich phases (blue data points in Fig.~4a), a trend that is
opposite to the converging tie lines from the dilute, protein-depleted 
phase boundary the condensed, protein-rich phase boundary in Fig.~2c.
This information is of critical importance, despite noted
experimental uncertainties in measuring concentrations, because
if the stabilizing interactions in the SynGAP/PSD-95 coacervate 
are only those between $\Syn_3\PSD_2$ complexes, the relative concentrations
of $\Syn$ and $\PSD$ in the condensed phase should not change or should
not vary much even when experimental uncertainties are taken into account.
As discussed above, the converging tie lines and the L-shape phase
boundary in Fig.~2c are direct consequences of the premise, based on
the scenario illustrated in Fig.~2b, that the only LLPS-driving favorable 
interactions are those among stoichiometric $\Syn_3\PSD_2$ complexes. 
In that hypothetical scenario, tie lines would have negative as well 
as positive slopes (positive slopes when overall
$\Syn$ and $\PSD$ concentrations are relatively low, 
negative slopes when either $\Syn$ or $\PSD$ concentration is high, see
Fig.~2c).
In contrast, the slopes of the experimental tie lines in Fig.~4a
are all positive, spanning an approximate 
range from $0.24$ to $9.1$.
It stands to reason, therefore, based on the fact that the trend inferred 
from experimental data in Fig.~4a is drastically different from 
that in Fig.~2c, it implies that favorable interactions 
among stoichiometric $\Syn_3\PSD_2$ complexes are not the only (or not 
at all) the LLPS-driving interactions in the assembly of the 
SynGAP/PSD-95 coacervate.
\\

%%%%%%%%%%%%%%%%%%%%%%%%%%%%%%%%%%%%%%%%%%%%%%%%%%%%%%%%%%%%%%%%%%%%%%%%%%%%%

\noindent
{\bf {\hbox{Theoretical analysis of experimental data suggests}} that 
non-stoichiometric auxilliary interactions are involved in
the assembly of SynGAP/PSD-95 coacervates}

To rationalize the SynGAP/PSD-95 phase properties in Fig.~4, it is useful
to note that the trend of diverging tie lines in Fig.~4a is similar to
that predicted by simple FH model for certain interaction
strengths, such as that reported in Fig.~6a of ref.~\cite{Lin2017c}.
Accordingly, we now apply FH approaches to assess alternate scenarios 
of SynGAP/PSD-95 coacervation (Fig.~5).
Because SynGAP/PSD-95 LLPS must involve interaction between $\Syn$
and $\PSD$ ($\Syn$ or $\PSD$ does not individually phase separate),
one may first consider a simple FH model with free energy
in units of $\kB TM$ given by
\begin{equation}
\begin{aligned}
f_1(\phi_{\Syn_3},\phi_{\PSD}) = 
& \; \phi_{\Syn_3}\ln \phi_{\Syn_3} + \phi_{\PSD}\ln \phi_{\PSD}  \\
	& + (1-\phi_{\Syn_3}-\phi_{\PSD})\ln(1-\phi_{\Syn_3}-\phi_{\PSD}) \\
	& \quad -\chi_{\Syn\PSD}\phi_{\Syn_3}\phi_{\PSD}
\; ,
\end{aligned}
	\label{eq:FH_pure}
\end{equation}
where $\chi_{\Syn\PSD}$ is the favorable energetic (FH $\chi$) parameter
characterizing $\Syn$--$\PSD$ contacts as the only LLPS-driving
interactions in this scenario (Fig.~5a). An example of the binary
phase diagrams predicted by this model is provided in Fig.~5c.
In contrast to the converging tie lines in Fig.~2c,
the tie lines in Fig.~5c exhibit some degree of divergence from the
protein-depleted to the protein-rich phase boundary, but the divergence
is not as prominent as that in Fig.~4a. Recognizing that the simple
free energy function $f_1(\phi_{\Syn_3},\phi_{\PSD})$ in Eq.~\ref{eq:FH_pure}
does not provide any bias toward formation of $\Syn_3\PSD_2$ complexes although
such complexes are observed experimentally in dilute solution \cite{Zeng2016},
it is reasonable to test whether incorporating favorability toward
formation of $\Syn_3\PSD_2$ complexes in the model would result in predicted
phase diagrams that conform better with the experimental trend.
We do so by making the following modifications to the free energy
$f(\phi_{\Syn_3},\phi_{\PSD})$ in Eq.~\ref{eq:final} (which already
contains favorable energy for $\Syn_3\PSD_2$ formation):
\begin{subequations}
	\label{eq:2hypo}
\begin{align}
\chi \psib^2 \; {\rm [Eq.~\ref{eq:final}]} 
&
\rightarrow \chi_{\Syn\PSD} \phi_{\PSD} \phi_{\Syn_3}
+ \frac {1}{2}[\chi_{\Syn\Syn} \phi_{\Syn}^2 + \chi_{\PSD\PSD} \phi_{\PSD}^2] 
\; , \\
\frac{\psib}{3}\Kdr^0 & = \psib_{\Syn_3}\psib_{\PSD}^2 e^{6\psib} 
{\rm [Eq.~\ref{eq:2step_Kd}]} \rightarrow \psib_{\Syn_3}\psib_{\PSD}^2 \; ,
\end{align}
\end{subequations}
wherein Eq.~\ref{eq:2hypo}a stipulates that individual $\Syn_3$ trimers
(not complexed with $\PSD$ in the manner of the $\Syn_3\PSD_2$ complex) 
and individual $\PSD$ molecules may participate in LLPS-driving
interactions and/or that higher-order SynGAP/PSD-95 complexes in addition
to $\Syn_3\PSD_2$ are present when protein concentrations are high and 
these higher-order complexes participate also in LLPS-driving interactiions
because the two scenarios are accounted for by the same highly coarse-grained, 
mean-field formulation 
in Eq.~\ref{eq:2hypo}a (Fig.~5b). Moreover, unlike the stoichiometric 
complex-driven LLPS model free energy in Eq.~\ref{eq:final}, 
by dropping the $\exp(6\psib)$ factor in Eq.~\ref{eq:2step_Kd}, 
Eq.~\ref{eq:2hypo}b signifies that $\Syn_3\PSD_2$ formation is
not coupled to the LLPS-driving interactions provided by the FH $\chi$ 
parameters (right hand side of Eq.~\ref{eq:2hypo}a) in the modified 
formulation described by Eq.~\ref{eq:2hypo}.
The modified formulation describes a system in which
$\Syn_3\PSD_2$ complexes are formed at low protein concentrations, 
whereas phase separation at higher protein concentrations is 
underpinned by a multitude of different $\Syn_3$ and $\PSD$ interactions.

As exemplified by the two example phase diagrams in Fig.~5d and e, the 
trend of tie-line divergence predicted by this modified formulation is more
akin to the experimental results in Fig.~4a than that predicted by the
simple FH model in Fig.~5c, 
although
the differences between the predicted tie-line patterns among Fig.~5c--e
are not dramatic.  Between the two models in Fig.~5d and e with
the same strongly favorable $\Syn$--$\PSD$ interactions,
it is instructive to note that the model in Fig.~5e with
favorable $\Syn$--$\Syn$ and $\PSD$--$\PSD$ interactions (consistent
with experiment, these interactions are
chosen to be insufficiently strong by themselves for $\Syn$ or $\PSD$
to phase separate individually)
exhibits more tie-line divergence and thus captures slightly better 
the tie-line trend in Fig.~4a than the model in Fig.~5d with only 
favorable $\Syn$--$\PSD$ interactions.
In particular,  the range of 
tie-line slopes,
$([\Syn^{\rm cond}]-[\Syn^{\rm dil}])/([\PSD^{\rm cond}]-[\PSD^{\rm dil}])$,
for the experimental results in Fig.~4a (slope $\approx 0.24$--$9.1$ as
mentioned above) are quite similar to that for the theoretical results 
in Fig.~5e (slope $\approx 0.28$--$12.0$, calculated by multiplying each 
volume-fraction slope by a factor of 3). This model feature
suggests that auxilliary $\Syn$--$\Syn$, $\PSD$--$\PSD$, 
and $\Syn$--$\PSD$ interactions in addition to those associated with 
$\Syn_3\PSD_2$ assembly, rather than auxilliary
$\Syn$--$\PSD$ interactions alone, 
are likely in play in the phase separation of the experimental SynGAP/PSD-95 
system as well. 
Nonetheless, the similarity of the tie lines
in Fig.~5c--e and the robustness of their overall pattern of divergence
indicate that the experimentally observed phase properties are
largely underpinned by auxilliary favorable $\Syn$--$\PSD$ interactions.
\\

%%%%%%%%%%%%%%%%%%%%%%%%%%%%%%%%%%%%%%%%%%%%%%%%%%%%%%%%%%%%%%%%%%%%%%%%%%%%%

\noindent
{\bf CONCLUSIONS}

%%%%%%%%%%%%%%%%%%

It has become increasing apparent that the assembly and disassembly of 
biomolecular condensates are driven by stochastic, ``fuzzy'' interactions 
involving intrinsically disordered protein regions
(see, e.g., refs.~\cite{Cliff2009,NatPhys2015,Nott2015,Kim2019})
as well as structurally specific interactions, including
stoichiometric binding of folded domains in some instances (see, e.g., 
refs.~\cite{Shorter2018,McKnightRev2018,Zeng2018a,SandersCliff2020,GITPIX}).
Deciphering how these interactions impart molecular recognition and
how they may act in concert to achieve function is of fundamental importance 
to molecular biology~\cite{ShorterRev2021,SchmitTiBS2021}, 
as has been emphasized in a recent study of  
the critical role of specific binding of the GIT1 and $\beta$-Pix GTPase 
regulatory enzymes in their complex coacervation~\cite{GITPIX}.
At a smaller length scale, a similar synergy between structurally specific 
interactions and stochastic, dynamic multivalent interactions has previously 
been indicated in the functional binding of IDPs in discrete, binary fuzzy 
complexes~\cite{Borg2007,Mittag2010}. In this context, the formation of 
the stoichiometric SynGAP/PSD-95 complex $\Syn_3\PSD_2$ in dilute 
solution \cite{Zeng2016} raises interesting questions regarding the
role of these stoichiometric complexes in SynGAP/PSD-95 LLPS. Intuitively,
a possible scenario would be that LLPS is driven solely by favorable 
interactions among $\Syn_3\PSD_2$ complexes. Theoretical considerations
reveal that this scenario implies a peculiar form of binary phase
diagrams with an L-shape phase boundary and converging tie lines. However,
subsequent experiments indicate that the LLPS properties of SynGAP/PSD-95
do not conform to this hypothetical scenario. Instead, our
pattern of phase behavior combined theoretical/experimental approach
suggests strongly that auxilliary interactions beside those responsible 
for the formation of $\Syn_3\PSD_2$ are also involved in stabilizing the
SynGAP/PSD-95 condensate. Our analysis has thus paved the way for 
investigating the nature of these auxilliary interactions. 
It appears that such LLPS-driving auxilliary interactions are readily
available for certain folded proteins, as in the cases
of lysozyme and crystallin~\cite{RolandRev}. It would be instructive to
compare SynGAP/PSD-95 with these simpler systems and explore also a 
possible involvement of disordered regions~\cite{Cliff-JCP2021}
in the formation of SynGAP/PSD-95 condensates. For instance,
based on the interface between the third PDZ 
and SH3-GK domains in PSD-95~\cite{Lee2013}, the contacts leading to 
higher order structures among PSD-95 molecules likely involve both 
hydrophobic and ionic interactions.  Moreover, bioinformatics 
considerations~\cite{IUPred3} suggest that a stretch of chain sequence
enriched in arginines, lysines, and glutamic acids 
between the coiled-coil and PBM domains of SynGAP
is likely disordered. Such IDRs are potential participants
in LLPS-driving interactions as well.
In this regard,
recently employed techniques for characterizing the roles of 
electrostatic and non-ionic interactions in biomolecular 
condensates~\cite{Rosana-NatComm2021,LEK-PNAS2021} would be useful.  
In any event, having 
demonstrated the capability of the relatively simple formulation developed 
here to gain fundamental insights into the mechanism of complex coacervation
of SynGAP and PSD-95, we hope that our methodology will be useful 
for analyzing the role of stoichiometric complexes in the assembly of 
other biomolecular condensates as well.
\\

\noindent
{\large\bf Author Contributions}\\
Y.-H.L., M.Z., H.S.C. designed research, 
Y.-H.L., H.W., B.J. performed research,
Y.-H.L., H.W. analyzed data, and
Y.-H.L., H.W., M.Z., H.S.C. wrote the article.
\\

\noindent
{\large\bf Acknowledgements}\\
We thank Julie Forman-Kay for helpful discussions.
We also thank three anonymous referees, whose constructive comments have
helped improve the clarity of our presentation.
This work was supported by Canadian Institutes of Health Research grant
NJT-155930 and Natural Sciences and Engineering Research
Council of Canada Discovery grant RGPIN-2018-04351 to H.S.C. 
and National Key R \& D Program of China grant 2019YFA0508402 and
RGC of Hong Kong grant AoE-M09-12 to M.Z. as well as computational 
resources provided generously by Compute/Calcul Canada.

\bibliographystyle{biophysj}
\bibliography{syn-pdz_HSC_S}

%$\null$\\ $\null$\\ $\null$\\ 
\onecolumngrid

\begin{figure}[ht]
\centerline{\LARGE\bf Figures} 
%$\null$\\ $\null$\\ 
{\includegraphics[height=95mm,angle=0]{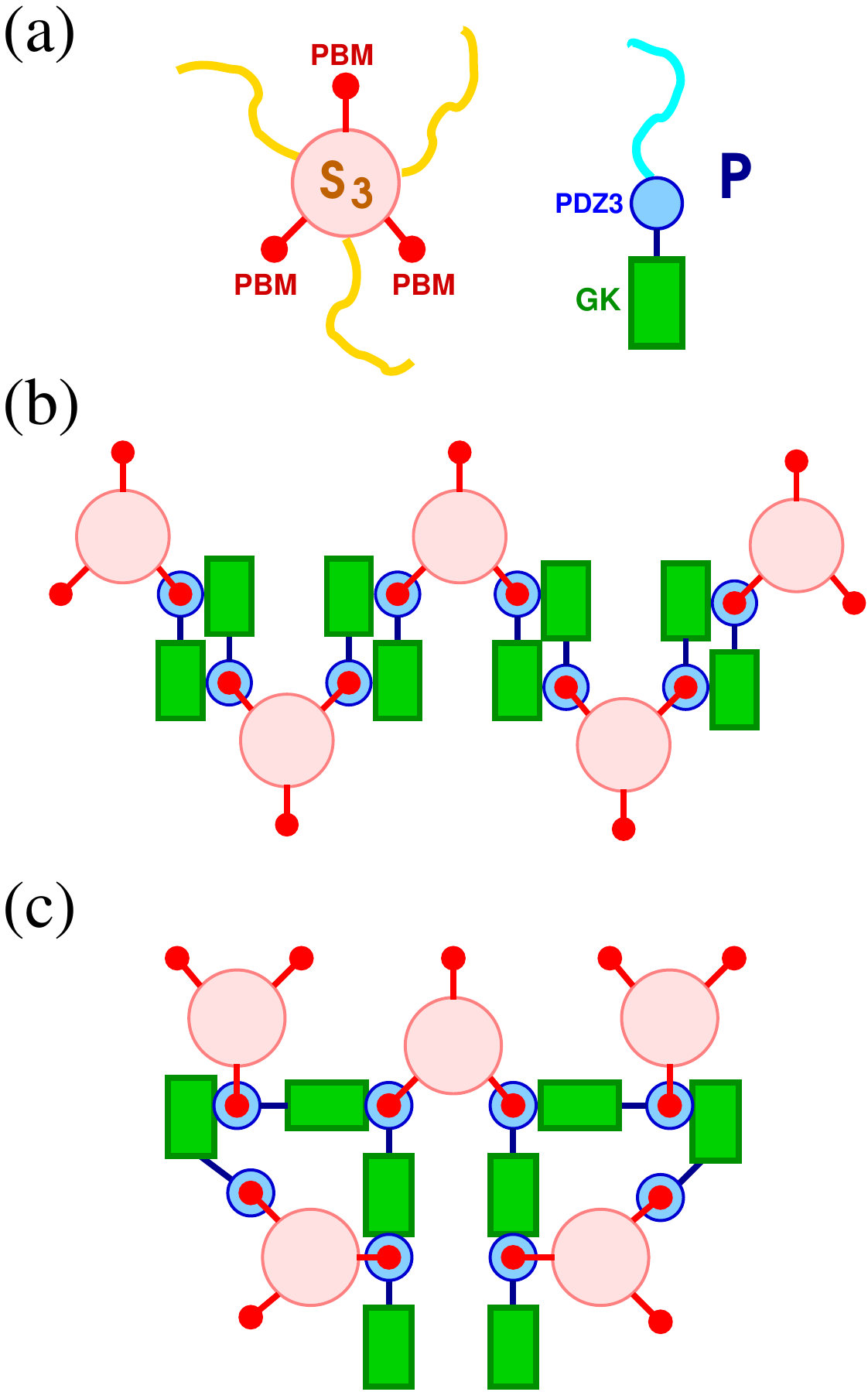}}
\caption{Schematics of hypothetical multiple SynGAP/PSD-95 configurations
that are based upon interactions among stoichiometric $\Syn_3\PSD_2$ complexes.
(a) Caricatures of the SynGAP trimer ($\Syn_3$, left) and PSD-95 
($\PSD$, right). For $\Syn_3$, the pink circle represents the SynGAP 
coiled-coil (CC) trimer, the three red circles represent the PDZ binding 
domains (PBMs) at the C-termini of the three $\Syn$ chains, and the three 
yellow chains represent the rest of each of the three $\Syn$ sequence 
N-terminal to the CC domain.
For $\PSD$, the blue circle represents the three PZD domains and
the green rectangle represents the C-terminal guanylate kinase (GK) domain.
The SH3 domain sequentially situated between these two domains is not
depicted explicitly. The blue curve represents the part of $\PSD$ N-terminal
to the PZD domains. Domain organizations of SynGAP and PSD-95 are described
in more detail in Figs.~1 and 2 of Ref.~\cite{Zeng2016}.
(b) $\Syn_3\PSD_2$ multimerization by connecting each $\Syn_3\PSD_2$ to 
two other $\Syn_3\PSD_2$s results in a linear chain of $\Syn_3\PSD_2$ 
complexes. This scenario corresponds to that in Fig.~7 of 
Ref.~\cite{Zeng2018b}. As described in this reference, the binding 
between an $\Syn_3$ with two $\PSD$ is mediated by PDZ-PBM contacts
(shown by concentric red and blue circles);
favorable interaction between two $\Syn_3\PSD_2$s is then effectuated
by binding of a PDZ/PBM of one $\Syn_3\PSD_2$ to the GK (and SH3) of another
$\Syn_3\PSD_2$ (indicated by a contact between a green rectangle with
a set of concentric red and blue circles). 
(c) $\Syn_3\PSD_2$ multimerization by connecting each $\Syn_3\PSD_2$ 
to three other $\Syn_3\PSD_2$s results in a network of $\Syn_3\PSD_2$ 
complexes. The notation for binding interactions is the same
as that in (b).
}
\label{fig1}
\end{figure}

\vfill\eject

\begin{figure}[ht]
{\includegraphics[height=55mm,angle=0]{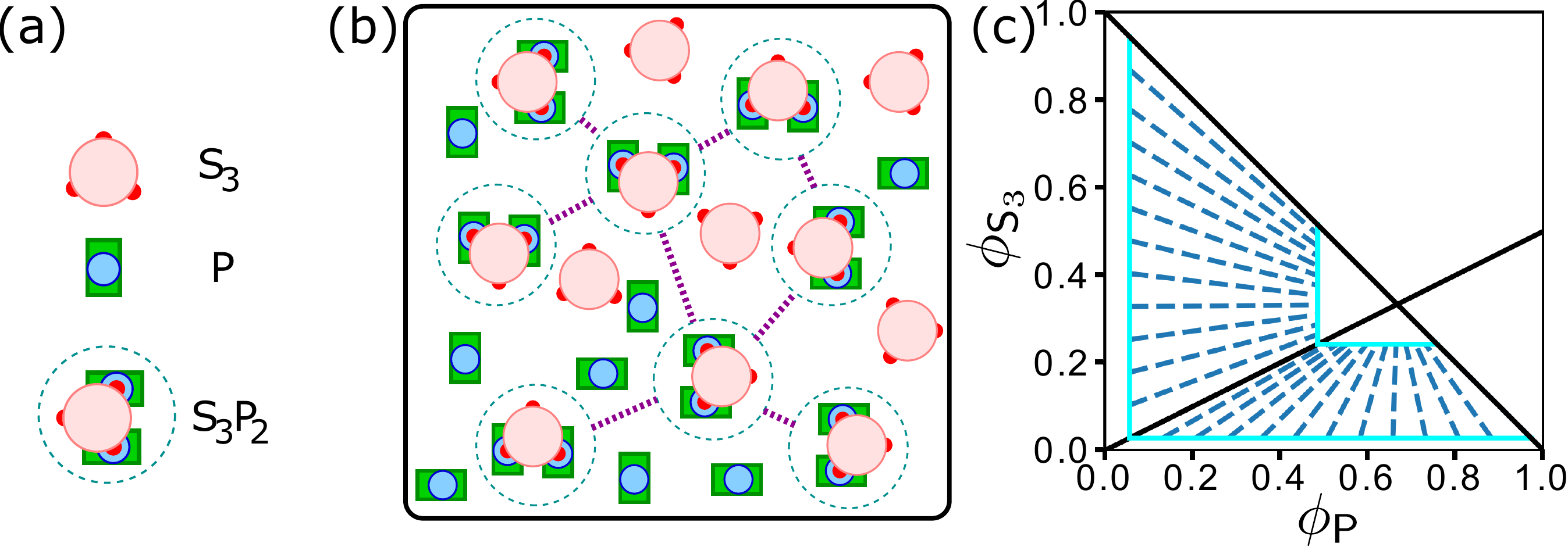}}
\caption{Prediction of a mean-field theory that assumes SynGAP/PSD-95 complex 
coacervation is driven soley by favorable interactions among stoichiometric 
$\Syn_3\PSD_2$ complexes (Eq.~\ref{eq:final}).
(a) Simplified schematic representation of $\Syn_3$, $\PSD$ and $\Syn_3\PSD_2$. 
The three $\Syn_3$ PBMs and their connecting chain 
segments to the CC trimer (large pink circle, see Fig.~1a) are now shown as
small red circles, the PDZ3-SH3-GK domains of $\PSD$ are now shown
as a blue circle encased in a green rectangle, and the yellow and blue 
N-terminal chain segments in Fig.~1a are not depicted explicitly.
Each $\Syn_3\PSD_2$ is enclosed by a dashed circle to underscore its role
as a unit of phase-separation-driving interaction in the mean-field theory.
(b) Schematic picture of a hypothetical SynGAP/PSD-95 condensed phase. 
The formulation in Eq.~\ref{eq:final} 
stipulates that while unbound S$_3$ and P may be
present in the condensed phase, phase separation is only driven by
interactions between units of S$_3$P$_2$ (magenta dashed lines).
These inter-unit favorable interactions may include binding of a PDZ/PBM 
of one $\Syn_3\PSD_2$ to the SH3-GK of another $\Syn_3\PSD_2$ as envisioned 
in Fig.~1b and c. 
(c) Predicted phase diagram for this hypothetical scenario based upon
Eq.~\ref{eq:final}, using ${\cal K}^0 = 10^{-10}$ and $\chi=1.5$ as 
illustration. The boundary of binodal phase separation is shown by the 
light blue lines. Each of the dashed dark blue lines is a tie line 
indicating a pair of coexisting phases on the boundary of the phase-separated
region. The overall volume constraint of $\phi_{\Syn_3}+\phi_{\PSD}\leq 1$ 
is marked by the solid black line with slope $=-1$ for 
$\phi_{\Syn_3}+\phi_{\PSD}=1$, whereas the $\phi_{\Syn_3}: \phi_{\PSD} = 1:2$ 
stoichiometric ratio of the $\Syn_3\PSD_2$ complex is indicated by the 
black solid line with slope $=1/2$.
}
\label{fig2}
\end{figure}

\vfill\eject

\begin{figure}[t]
{\includegraphics[height=105mm,angle=0]{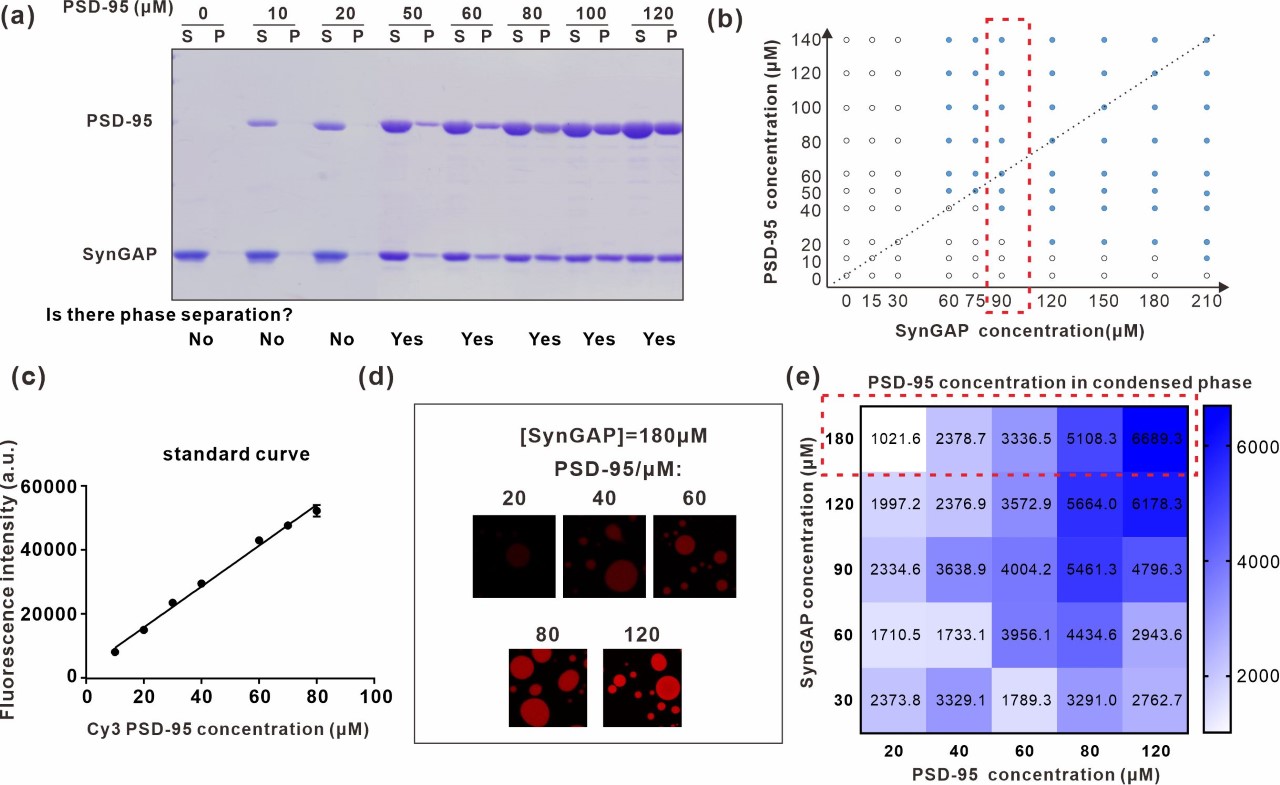}}
\caption{Experimental measurements of SynGAP/PSD-95 phase separation.
(a) Sedimentation assay showing the dilute phase (marked by ``S'' at the
top of the columns) and the condensed phase (marked by ``P'' at the top
of the columns) distribution of PSD-95 and SynGAP mixed at different 
concentrations of PSD-95 and a fixed concentration of SynGAP at 90 $\mu$M.
(b) Phase diagram of PSD-95 and SynGAP condensates. The hollow circles 
in phase diagram indicate no phase separation and solid circles 
represent condensed phase formation.
The results enclosed in the dashed red box correspond to those from (a).
The inclined dashed line indicate the dilute-solution
3:2 ratio for [SynGAP]:[PSD-95].
(c) Standard curve of Cy3-labeled PSD-95 obtained from dilute solutions. 
(d) Confocal images showing an indicative confocal slice of condensed 
droplets of each concentration of Cy3-labeled PSD-95 mixed with unlabeled 
SynGAP at a fixed concentration at 180 $\mu$M. 
(e) Heat map plot showing the PSD-95 concentrations in condensed phase at 
different combinations of PSD-95 and SynGAP concentrations. Each number 
represents the PSD-95 concentration ($\mu$M) in the condensed phase.
The data enclosed in the dashed red box correspond to those from (d).
}
\label{fig3}
\end{figure}

\vfill\eject

\begin{figure}[t]
{\includegraphics[height=70mm,angle=0]{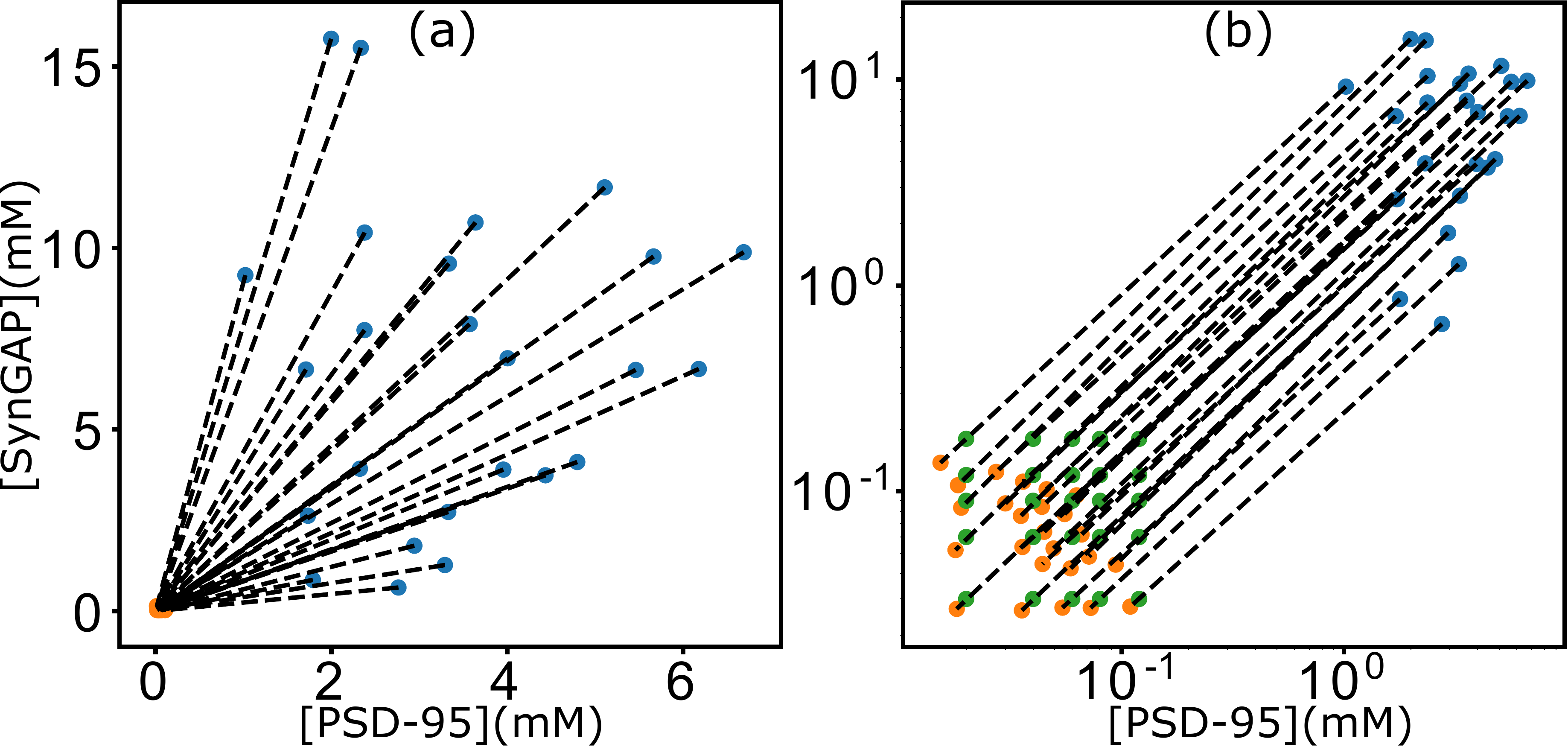}}
\caption{Experimental trend of SynGAP/PSD-95 coexisting phases. Shown
data are inferred from confocal microscopy and centrifigation measurements
as described in the text. Initial (overall) concentrations
$([\PSD^0], [\Syn^0])$ are plotted as green data points. The phase-separated
dilute-phase $([\PSD^{\rm dil}],[\Syn^{\rm dil}])$ are plotted as 
orange data points, whereas the condensed-phase 
$([\PSD^{\rm cond}],[\Syn^{\rm cond}])$ are plotted as blue data points.
Dashed lines are tie lines connecting coexisting phases.
Results are shown in linear (a) as well as log-log (b) scales for clarity, as
the green data points are not visible in the linear plot in (a) because 
they are very close to the orange data points. 
Note that because the 
dilute-phase $([\PSD^{\rm dil}],[\Syn^{\rm dil}])$ concentrations 
(orange data points) are all very close
to the origin, the slopes of tie lines in the log-log plot in (b) are
all approximately equal to one. The diversity of the actual tie-line slopes
as seen in (a) is manifested by the offsets of the 
log-log tie lines (different intercepts) in (b).
}
\label{fig4}
\end{figure}

\vfill\eject

\begin{figure}[t]
{\includegraphics[height=120mm,angle=0]{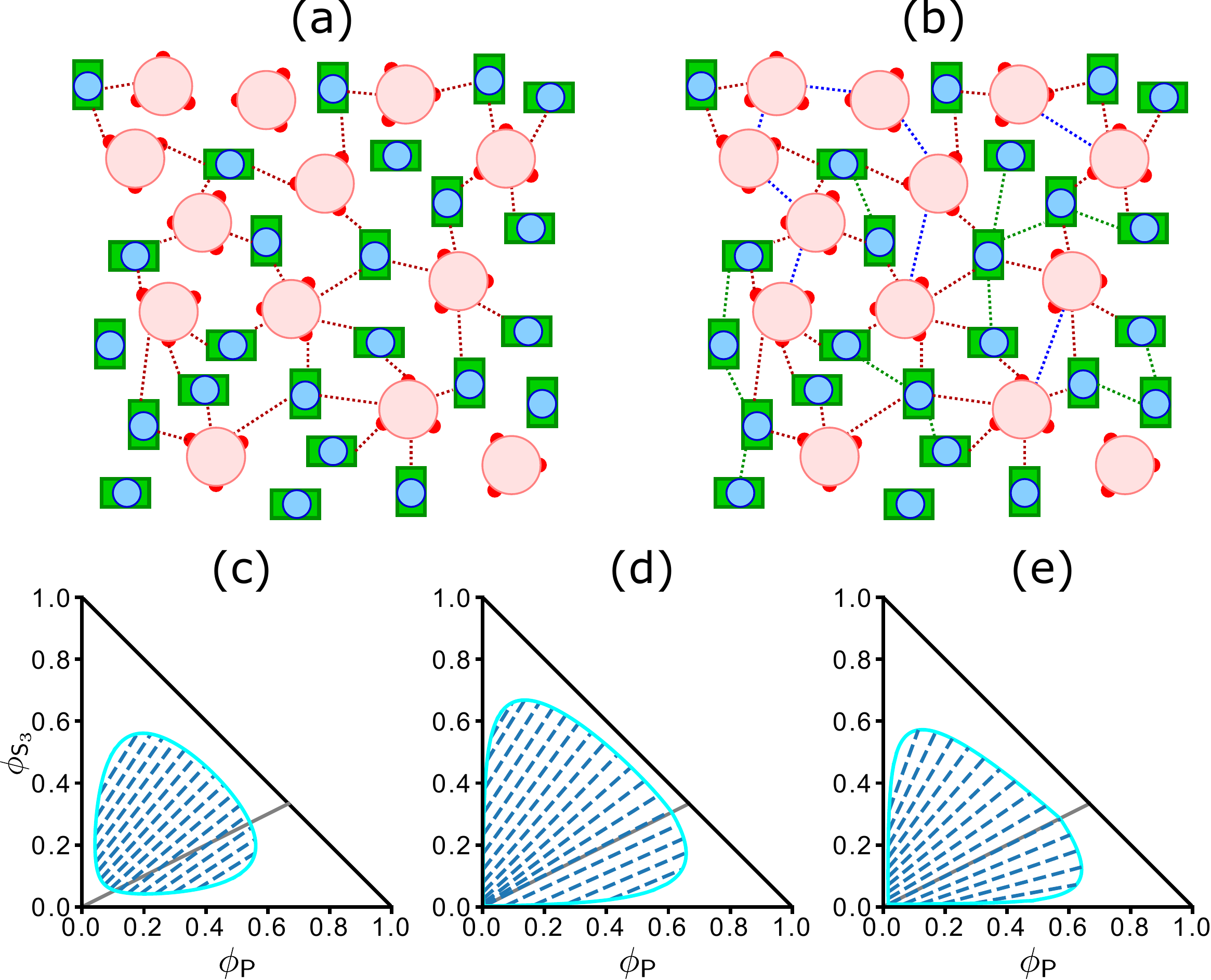}}
\caption{Alternate scenarios of SynGAP/PSD-95 phase separation with 
auxilliary interactions beyond those underpinning the assembly of
stoichiometric $\Syn_3\PSD_2$ in dilute solution. The schematic 
representations of $\Syn_3$ and $\PSD$ here are the same as those in Fig.~2a.
(a) Favorable LLPS-driving interactions are envisioned to be restricted
to those between $\Syn_3$ and $\PSD$ (red dashed lines) 
as modeled by the $\chi_{\Syn\PSD}$ term in Eq.~\ref{eq:FH_pure} for
a simple FH model. 
(b) LLPS is envisioned to be driven also by favorable interactions 
among $\Syn_3$ (blue dashed lines) and among $\PSD$ (green dashed lines) as
modeled by the $\chi_{\Syn\Syn}$ and $\chi_{\PSD\PSD}$ terms in 
Eq.~\ref{eq:2hypo}a. The bias afforded by this model toward formation 
of $\Syn_3\PSD_2$ complexes in the dilute phase is not exhibited in 
this schematic depiction of the condensed phase.
(c)--(e) Mean-field FH theory predictions for 
SynGAP/PSD-95 phase behaviors in the alternate scenario in (a)
[(c) and (d)] and the alternate scenario in (b) [(e)].
The black solid lines with slopes $=1/2$ and $-1$, 
the dashed tie lines, and the light blue phase boundaries in the 
three phase digrams carry the same meanings as those in Fig.~2c.
The FH parameters are:
(c) $\chi_{\Syn\PSD}=10.0$ in Eq.~\ref{eq:FH_pure},
(d) $\Kdr^0=10^{-10}$, $\chi_{\Syn\PSD}=10.0$ 
and $\chi_{\Syn\Syn}=\chi_{\PSD\PSD}=0$ 
in Eq.~\ref{eq:2hypo}, and
(e) $\Kdr^0=10^{-10}$, 
$\chi_{\Syn\PSD}=4.0$ and $\chi_{\Syn\Syn}=\chi_{\PSD\PSD}=3.5$
in Eq.~\ref{eq:2hypo}.
}
\label{fig5}
\end{figure}

\vfill\eject

\vfill\eject

\end{document}